\documentclass[]{eptcs}
\providecommand{\event}{AREA 2023}

\usepackage{iftex}
\usepackage{graphicx}
\usepackage{wrapfig}
\usepackage{subcaption}
\usepackage{listings}
\usepackage{xcolor}
\hypersetup{hidelinks} 

\ifpdf
  \usepackage{underscore}         
  \usepackage[T1]{fontenc}        
\else
  \usepackage{breakurl}           
\fi

\title{Safe and Robust Robot Behavior Planning via~Constraint~Programming}

\author{Jan Vermaelen \qquad\qquad\qquad\quad Tom Holvoet
\institute{imec-DistriNet, KU Leuven, Belgium}
\email{jan.vermaelen@cs.kuleuven.be \quad\qquad tom.holvoet@cs.kuleuven.be}
}
\def\titlerunning{Safe and Robust Planning}
\def\authorrunning{J. Vermaelen \& T. Holvoet}
\begin{document}
\maketitle

\begin{abstract}
The safe operation of an autonomous system is a complex endeavor, one pivotal element being its decision-making. 
Decision-making logic can formally be analyzed using model checking or other formal verification approaches.
Yet, the non-deterministic nature of realistic environments makes these approaches rather troublesome and often impractical.
Constraint-based planning approaches such as Tumato have been shown to be capable of generating policies for a system to reach a stated goal and abiding safety constraints, with guarantees of soundness and completeness by construction.
However, uncertain outcomes of actions in the environment are not explicitly modeled or accounted for, severely limiting the expressiveness of Tumato.

In this work, we extend Tumato with support for non-deterministic outcomes of actions.
Actions have a specific intended result yet can be modeled to have alternative outcomes that may realistically occur.
The adapted solver generates a policy that enables reaching the goals in a safe manner, even when alternative outcomes of actions occur.
Furthermore, we introduce a purely declarative way of defining safety in Tumato, increasing its expressiveness.
Finally, the addition of cost or duration values to actions enables the solver to restore safety when necessary, in the most preferred way.
\end{abstract} 

\section{Introduction}\label{introduction}
Autonomous robotic systems are becoming increasingly popular, both in industry and households.
The number and complexity of tasks they are expected to execute are expanding.
However, generating a plan providing both productive (goal-oriented) and safe behavior is far from trivial.
A plan must be constructed, given the actions that the robot can execute, the information about the environment, and the desired goals.
During planning, additional safety constraints have to be taken into account to generate a safe plan while trying to achieve the goals.
If such a safe and productive plan can be generated, it is sound by construction.

The main contribution of this paper is to support foreseeable non-deterministic transitions while guaranteeing safety when planning robot behavior.
For this purpose, we focus on constraint-based planning and build further upon \textbf{Tumato}, a planning framework by Hoang Tung Dinh et al.~\cite{DinhHoangTung2017Sacr}.
Firstly, we support specifying and accounting for foreseeable non-deterministic \textit{alternative effects} of actions rather than assuming a purely deterministic system.
Effects of actions in the real world are virtually never fully deterministic.
Non-determinism can arise from varying weights of payloads, various ground surfaces to navigate across, and small measuring and actuation errors, to name just a few.
Secondly, the extension supports specifying \textit{safety conditions} explicitly.
In the original version of Tumato, to obtain a similar result, one has to consider all state-action combinations that could lead to unsafe states separately, which is error-prone and inflexible when the specification of the system evolves.
Furthermore, the generated behavior \textit{always} has to adhere to the safety conditions, even when alternative effects occur.

Since the environment, and hence the effects of actions, are most often not deterministic in practical robotic applications, the policy must be sufficiently robust to deal with this kind of uncertainty.
Ideally, all contingencies are taken into account.
A first step is to pursue a \textit{complete} policy.
For each state in which the system could be, the policy must provide the actions to execute next.
If the system unexpectedly arrives in an unintended state, the operation can continue.
We maintain this powerful feature of Tumato's original approach.
In a second step, we take the uncertainty into account by allowing the effects of actions to be modeled in a non-deterministic way.
In our approach, we assume that each action has one intended outcome.
We call this outcome the \textit{nominal} effect of the action.
Additionally, each action can have a number of \textit{alternative} effects.
These effects \textit{could} emerge instead of the nominal one, but they are not intentional.
For the goal-oriented aspect of planning, only the nominal effect is relevant.
Alternative effects are unintentional outcomes and can not reliably be used to achieve a goal.
However, when dealing with safety, also the alternative effects must be taken into account as there is a possibility they occur.

Due to external causes, the system might still end up in an \textit{unforeseen} state.
Although for such events safety can not be guaranteed (an external force might put the system in an unsafe state directly), the planner will make sure that the policy contains instructions on how to get back on (safe) track to the goal immediately.
If multiple such instructions are possible, the planner is capable of selecting the preferred one based on cost values assigned to actions.

The paper is structured as follows.
Section~\ref{Related work} discusses the related work to outline the necessary background and provides an overview of Tumato.
Section~\ref{casestudy} briefly introduces the use case of the robotic system used to illustrate the proposed extension.
Section~\ref{safetyanduncertainty} explains and motivates the approach.
Section~\ref{specification} elaborates and analyzes the extension of the specification using the robotic system from the case study.
Section~\ref{discussion} discusses the approach and its results.
Finally, Section~\ref{conclusion} draws conclusions.

\section{Background and Related Work}\label{Related work}
Well-thought-out behavior planning is essential for the safe operation of autonomous safety-critical robots.
Furthermore, practical systems often involve a degree of non-determinism that needs to be addressed.
In this section, we point to the related work necessary to provide a background for the explored planning approach.
However, all (fully observable) non-deterministic planning can be considered related.

Traditionally, the behavior of robots has been defined manually.
Finite State Machines (FSMs) are often used to represent the robot's behavior~\cite{FSMbsorcsa, Nguyen2013ROSC}.
However, FSMs are known not to cope well with the increasing complexity of the behavior.
It is non-trivial to manually specify sound and complete behavior for larger and more complex systems, and one has to rely on simulation and verification approaches to check whether the behavior effectively meets the requirements.
This problem can partly be solved by automatically generating the behavior based on a model of the system, along with a representation of the desired requirements.

Different specification languages and planners have been proposed.
For example, Linear Temporal Logic (LTL)~\cite{emerson1990temporal} is often used in robotics.
Techniques exist to use (fragments of) LTL to generate FSMs~\cite{maniatopoulos2016reactive, wongpiromsarn2013synthesis}, or to compile them into PDDL~\cite{camacho2016non}.
Since LTL can take all contingencies into account, the generated behavior will be sound and complete, a property we also value.
Further, modeling state-based safety conditions explicitly in LTL should require limited effort.
Two other examples are Temporal Action Logic (TAL)~\cite{doherty1998temporal, doherty2001talplanner} and the previously mentioned, more generic Planning Domain Definition Language (PDDL)~\cite{gerevini2009deterministic, karpas2020automated}.
Both TAL and PDDL generally rely on replanning at run-time to cope with contingencies.
They do not guarantee completeness of the behavior since the replanning could fail due to an unrealizable specification.
This lack of completeness would only be detected at run-time.
Similarly, in a more practical context, probabilistic planners can be used within the ROSPlan~\cite{cashmore2015rosplan} framework directly~\cite{canal2019probabilistic}.

To a certain extent, robustness can be obtained by explicitly dealing with non-determinism, as covered in the book \textit{Automated Planning and Acting} by Malik Ghallab et al.~\cite{AutomatedPlanningandActing}.
The planning can freely make use of the non-deterministic effects of actions to reach the goal.
Unlike this approach, we opt to define one (intended) nominal effect for each action while recognizing alternative (less likely and less desired) outcomes.
This is more closely related to practical behavior planning problems.
Note that, in the non-deterministic context, the definition of a \textit{Safe Solution} is a policy in which the goal is reachable from the initial state~\cite{AutomatedPlanningandActing}.
This definition is different from the additional safety constraints that we are imposing on the system to reach the goal \textit{in a safe manner}.

Tomas Geffner et al. introduce a SAT encoding for fully observable non-deterministic planning~\cite{geffner2018compact}.
A distinction is made between \textit{fair} and \textit{unfair} non-deterministic actions.
In our work, we focus on fair actions and do not consider adversarial actions or agents.
We do acknowledge that probabilistic effects of actions are difficult to estimate correctly, while they should not be considered fully non-deterministic either.

We have surveyed existing frameworks combining safe and robust planning before~\cite{vermaelen2020survey}.
The use of Markov Decision Processes~(MDPs)~\cite{puterman2014markov} for probabilistic planning and, to a smaller extent, Simple Temporal Networks~(STNs)~\cite{dechter1991temporal} for temporal scheduling have been explored.
Especially their extensions are able to explicitly guarantee safety while dealing with uncertainties.
In this work, avoiding the need for (often inherently imprecise) probabilistic values, we investigate and extend the promising constraint programming approach Tumato by Hoang Tung Dinh et~al.~\cite{DinhHoangTung2017Sacr}.

\paragraph{Tumato}\label{TumatoOld}
Hoang Tung Dinh et al.~\cite{DinhHoangTung2017Sacr} obtain sound and complete behavior via constraint programming.
As the specification of a system has to contain information about the environment, the actions, and the goals of the system, as well as a set of safety rules, it effectively combines classical planning (using states, actions, and goals) with constraint programming to enforce safety.

Constraint-based planning is achieved by (automatically) translating the entire specification into constraints.
Trivially, preconditions of actions constrain whether or not the action can be executed.
Furthermore, the effects of actions on the state of the system and whether or not an action is executed in the first place are modeled as constraints.
One Constraint Satisfaction Problem (CSP)~\cite{tsang1993foundations} is generated for each valid starting state of the system.
The set of CSPs yielding from the specification can be solved offline.
Tumato currently employs Choco-solver~\cite{prud2022choco}, yet we try to maintain transparency to the exact solver used.
The first execution step found by each individual CSP is selected as the set of actions for the corresponding state.
The final result is a mapping from every state to the actions that have to be executed in that state.
We will call this mapping the \textit{policy}.
If a solution exists, we consider the specification to be realizable.
Otherwise, we say the specification is unrealizable.
If a sound and complete policy exists, it will be found by the constraint solver.
The obtained policy can safely be used at run-time without requiring online re-planning.
For further details on the general approach, please refer to the original work on Tumato~\cite{DinhHoangTung2017Sacr}.

In the remainder of this section, we will give an overview of the specification of a model in Tumato.
The specification contains information on the state space of the system, the actions that can be executed, the relevant safety constraints, and the goals of the system.
Our work will extend this specification where necessary, as described in Section~\ref{specification}.

\paragraph{States}
A state vector is used to represent the states in which the system can reside.
This vector consists of a set of discrete state variables.
Each of the state variables represents one aspect of the state.
At run-time, the state variables are updated by a monitoring module.
This module is responsible for translating the sensor readings and other input to discrete values in the state variables.
An example of a state variable is the location of the robot.
A state variable \(S_{Location}\) can represent the different discrete locations at which the robot can reside.
For example, \{\textit{corridor, charger, workstation\_1, workstation\_2}\} includes a common corridor, a charging station for the robot, and two interactable workstations. Additional workstations and locations can be added as needed.
Alternatively, when more detailed location information is required, a more fine-grained location discretization can be utilized.

\paragraph{Actions}\label{TumatoActions}
Actions represent elemental behavior.
They are the smallest unit of execution that we consider during planning.
As an example, both \textit{move\_one\_cell\_forward} and \textit{move\_to\_charging\_station} are possible actions, yet they relate to planning at different abstraction levels.
Actions can be specified to have \textit{preconditions} that must hold before they can be executed.
Actions also can be specified to make use of certain \textit{resources}.
The purpose of allocating resources is to prevent two actions from being simultaneously executed if they make use of at least one resource in common.
Finally, actions usually have an \textit{effect} on the state.
After executing an action, which can take an arbitrary amount of time, the state has changed according to that effect.
An example of a simple action is \textbf{move_to_workstation_1}, which:
\begin{itemize}
    \item controls one resource: \textit{motors},
    \item has one precondition: \(S_{Location} = corridor\),
    \item has the effect: \(S_{Location} = workstation\_1\).
\end{itemize}

\paragraph{Reaction Rules}\label{TumatoRRules}
As their name indicates, reaction rules can be used to specify reactive behavior.
They are logical rules on the current state and the executed actions.
If a certain condition holds on the current state, either a specific action has to be executed or is not allowed to be executed.
For example, \textit{if the robot resides in the corridor with a workstation-sensitive actuator (for example, a conveyor belt) active, then it should deactivate that actuator (for example, by executing the action \textbf{stop\_conveyor})}.
Let \(S_{Conveyor}\) represent whether or not the conveyor is currently active, as \{\textit{on, off}\}.
\[(S_{Location} = corridor \wedge S_{Conveyor} = on) \Rightarrow Exec(\textbf{stop\_conveyor}).\]
We will discuss the use of reaction rules in more detail in Section~\ref{reactionrules}.

\paragraph{Goals}
The goals define conditions that have to be achieved by the system.
Because of the constraint-based approach, different formats of goals can be specified.
For this example, we focus on \textit{prioritized} and \textit{conditional} goals. A conditional goal is a goal that is active only when a specified condition is met.
Consider a state variable \(S_{Load}\) that can be \{\textit{loaded, free}\}.
\textit{If the robot is loaded with an item, the goal is to unload that item}~(\ref{deliver}) and hence, deliver it.
Analogously for picking up an item, a conditional goal is specified~(\ref{obtain}).
Priorities can indicate which conditional goals should be taken into account first.
Consider a state variable \(S_{Battery}\) that can be \{\textit{low, ok}\}.
A conditional goal can be responsible for recharging the battery when the battery level becomes \textit{low}~(\ref{battery}).\footnote{For a practical application, it is important that the monitoring module only updates \(S\_Battery\) to \textit{ok} when the battery has been sufficiently charged. An earlier update would result in (undesired) shorter operation cycles.}
This conditional goal should get priority over the transport goals (\ref{deliver}) and (\ref{obtain}).
Priorities are determined by the order in which the goals are specified.

\begin{equation}\label{battery}
    (S_{Battery} = low) \Rightarrow S^{Goal}_{Battery} = ok
\end{equation}
\begin{equation}\label{deliver}
    (S_{Load} = loaded) \Rightarrow S^{Goal}_{Load} = free
\end{equation}
\begin{equation}\label{obtain}
    (S_{Load} = free) \Rightarrow S^{Goal}_{Load} = loaded
\end{equation}

\section{Case Study}\label{casestudy}
\begin{wrapfigure}{R}{0.33\textwidth}
\vspace{-12mm}
\begin{center}
    \includegraphics[width=0.95\linewidth]{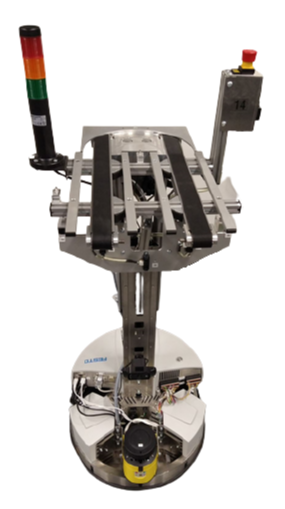}
    \caption{The AMR}\label{fig:AMR}
\end{center}
\vspace{-8mm}
\end{wrapfigure}

As hinted toward in the previous section, we consider a battery-powered Autonomous Mobile Robot (AMR) for validation purposes.
The AMR, shown in Figure~\ref{fig:AMR}, operates in an automated demo factory.
The workstations present in the factory are capable of executing different operations on small, standardized workpieces.
Adjacent workstations share a conveyor belt, yet not all workstations are connected.
The AMR is responsible for moving around workpieces in the factory.
We use this system as an example to paint out shortcomings of the current constraint programming planning approach and to illustrate how to mitigate these shortcomings.

The base of the AMR has three omnidirectional wheels (not visible in the picture), enabling the AMR to move in all directions as well as to turn in place.
Combined with information regarding the layout of the factory, the AMR can move to any location and obtain any orientation within the factory.
Only 2D movement and orientation are applicable.
The AMR can navigate between the different workstations and a charging station, as well as the \textit{corridor} connecting the different locations.
The AMR has a camera and LIDAR system for perception, along with a number of proximity sensors.
On top of the base, a looping conveyor belt is mounted at a fixed height.
This setup is used to dock to workstations and receive or deliver workpieces from or to workstations.
Infrared sensors attached near the conveyors help with the alignment when docking.
Furthermore, an emergency stop button is mounted on top, as well as a beacon for visual signaling.

One or more AMRs can be operating simultaneously on the factory floor alongside human agents.
The AMRs and humans are not physically separated, increasing the importance of generating safe behavior.
An AMR should complete its assigned transport without creating any unsafe situations.
As an additional hurdle, the effects of actions can unintentionally vary, even within a relatively controlled environment such as a factory.

\section{Uncertainty and Safety}\label{safetyanduncertainty}
As mentioned in Section~\ref{introduction}, we account for uncertainty by adding foreseeable alternative outcomes to actions.
Merely to illustrate the approach, this section introduces a grid-based 2D navigation planning example, although the approach can be more fully appreciated with examples from the case study, see Section~\ref{specification}.

\subsection{Uncertainty}\label{uncertainty}
\begin{figure}
\centering
\begin{subfigure}{.28\textwidth}
    \centering
    \includegraphics[width=.66\linewidth]{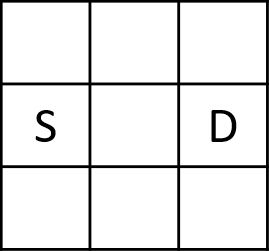}  
    \caption{Starting point \(S\) and destination \(D\)}
    \label{3x3grid}
\end{subfigure}
\hspace{2mm}
\begin{subfigure}{.28\textwidth}
    \centering
    \includegraphics[width=.66\linewidth]{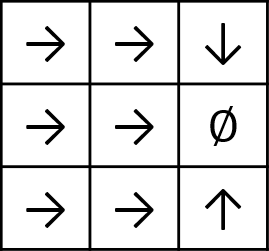}  
    \caption{Policy for a deterministic approach}
    \label{3x3detpolicy}
\end{subfigure}
\hspace{2mm}
\begin{subfigure}{.28\textwidth}
    \centering
    \includegraphics[width=.66\linewidth]{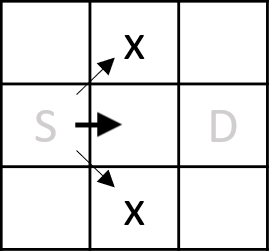}  
    \caption{\textit{Move\_right} in \(S\) with two alternative outcomes ($\mathsf{X}$)}
    \label{3x3nondet}
\end{subfigure}
\caption{Illustrative \(3 \times 3\) grid example}
\vspace{-5mm}
\end{figure}

Assume a \(3 \times 3\) grid, with starting point \(S\) and destination \(D\) as illustrated in Figure~\ref{3x3grid}.
We assign each \textit{cell} in the grid to a corresponding state.
Further, we define the actions \textit{move\_up, move\_down, move\_left}, and \textit{move\_right}.
In a deterministic setting, each action always moves the agent exactly one cell in the intended direction of that action.
In this example, the shortest plan from starting point \(S\) to destination \(D\) executes \textit{move\_right} twice.
An example of a complete policy for this planning problem is shown in Figure~\ref{3x3detpolicy}.

More realistically, however, in a non-deterministic planning setting, actions are not always successful.
Firstly, the action could have no effect.
For example, the agent did not move far enough to reach the new cell.
Since we are dealing with policy-based execution, the same action will be executed again, presumably leading to a new state eventually.
Note that this implicit assumption has to be taken into account when constructing the system.
If an action could fail indefinitely, for example, if the agent could have insufficient force to move up a hill, the action should be reconsidered.
Either the model is not adequate (the effect does not correspond to the real world), or the (physical) implementation of the action has to be adapted.

Secondly, deviations can occur.
When executing \textit{move\_right} at the starting point \(S\), two states (cells) could be reached unintentionally, as shown in Figure~\ref{3x3nondet}.
Such outcomes could either be observed after deploying and running the system for an extended period or be indicated by experts.
These outcomes are less likely than the desired and intentional (nominal) outcome, which is the middle cell in the \(3 \times 3\) grid.
Since these outcomes can occur, we should be aware of them when modeling the system, especially when dealing with safety.
However, as the alternative outcomes only occur sporadically, we can not determine accurate probabilistic values to use during planning.\footnote{If one is able to determine the probabilistic values for the effects accurately, using an MDP would be a better choice to obtain the policy.}
For the productive, goal-oriented aspect of planning, we do not take into account alternative outcomes.
The constraint solver will assume that the intended effects occur.
If, during execution, the system were to arrive in any unintentional state (possibly due to an alternative effect), the completeness of the policy will make sure that the operation can continue to reach the goal.
If from such a state, and by extension from \textit{any} state, the goal can not be reached no policy can be found.
The planner will then provide such states as feedback to the developer.

\paragraph{Note on non-determinism}
The presented approach, using nominal and alternative effects, can also be used to model true non-determinism in the system.
We illustrate this with the example of a coin flipped at run-time.
Two actions (one for each side of the coin) are modeled to eliminate the inherent bias toward the nominal outcome.
For each action, the nominal effects take care of one side of the coin while the alternative effects take care of the other side.
Regardless of the outcome of an individual toss, the policy ensures that the executed actions are safe, as we define next.

\subsection{Safety}
We extend the previous grid example to a \(5 \times 5\) grid.
In this example, the outer cells are considered unsafe.
These cells could be located next to stairs which the agent could fall off, or a wall it could run into.
Although simply defining unsafe states or conditions is a very straightforward way to define~(un)safety, it yields a powerful approach.
It provides a large improvement over manually identifying which state-action combinations are unsafe, as the original Tumato planner requires using reaction rules.

\begin{figure}
\centering
\begin{subfigure}{.48\textwidth}
    \centering
    \includegraphics[width=.6\linewidth]{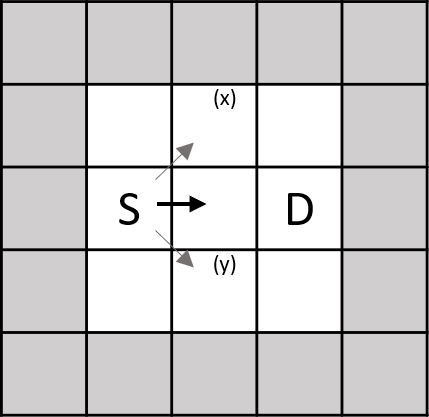}  
    \caption{Visualization of the three reachable states by \textit{move\_right} from S. A second \textit{move\_right} (not visualized) in either \textit{x}~or~\textit{y} could lead to unsafe (gray) states.}
    \label{5x5action}
\end{subfigure}
\hspace{2mm}
\begin{subfigure}{.48\textwidth}
    \centering
    \includegraphics[width=.6\linewidth]{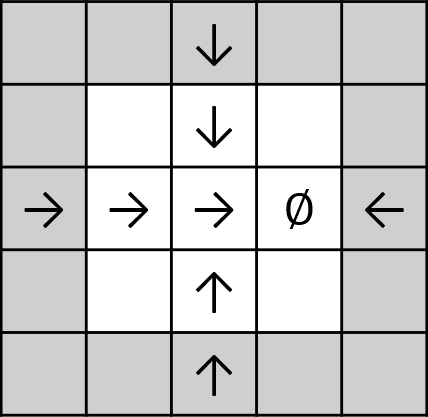}  
    \caption{A guaranteed safe policy. This policy should \textbf{not} be considered complete unless it is accompanied by the necessary assumptions.}
    \label{5x5safepolicy}
\end{subfigure}
\caption{Illustrative \(5 \times 5\) grid example}
\vspace{-5mm}
\end{figure}

A \textit{move\_right} at the starting point \(S\) will never lead to an unsafe state, as shown in Figure~\ref{5x5action}.
This action will be allowed to be used during planning.
In turn, in both of the unintentionally reachable states (separately denoted as \textit{x} and \textit{y} in the figure), a similar \textit{move\_right} will not be allowed, as alternative outcomes could lead the agent into an unsafe state.
A safe policy is shown in Figure~\ref{5x5safepolicy}.
Note that many cells remain empty.
Those cells do not allow any safe actions in our current example.
To obtain this policy, the specification would explicitly have to contain a number of assumptions to exclude those cells from the valid state space.
If those assumptions are not added, no result will be obtained since no complete policy exists.
In this paper, we will not look further into assumptions or how they are modeled.

\paragraph{Note on restoring safety}\label{restoringSafety}
The approach mentioned above does not distinguish between starting from a safe or unsafe state: the next state is guaranteed to be safe.
Hence, for unsafe states, a solution will only be found when safety can be restored within one step.
In addition, when multiple actions are executed, the planner is aware that only the effects of \textit{one} of those actions will take place \textit{first}.
As a result, every possible outcome of every action selected for execution must be safe.
Furthermore, since actions can be assigned a cost value (see Section~\ref{NewActions}), the planner can choose the set of actions that restores safety in the most preferred way.

\section{Specification}\label{specification}
In this section, we elaborate on the modifications made to the specification introduced in Section~\ref{TumatoOld}.
Our contribution lies in the inclusion of alternative effects and costs for actions, as well as the addition of state rules.
We use the AMR described in Section~\ref{casestudy} as an example for constructing a formal specification.
For the complete specification, formatted as supported by Tumato, please refer to Appendix~\ref{appendix}.

\subsection{Actions}\label{NewActions}
Unlike in the original Tumato framework, we define two kinds of effects.
The \textit{nominal effects} represent the desired and intended outcome of an action, whereas the \textit{alternative effects} correspond to unintended outcomes that \textit{could} occur instead of the nominal one, but are less likely to.
To provide a comprehensive understanding of both kinds of effects, we present a two-fold perspective.
In the context of productive (goal-oriented) planning, only the nominal effects are considered, as mentioned in Section~\ref{uncertainty}.
Undesired outcomes are deliberately excluded.
However, when ensuring safe behavior, \textit{all} effects should be considered.
Each possible outcome of every executed action must be safe.
This two-fold perspective enables the appreciation of both kinds of effects within their respective formal contexts.

Further, we extend the specifications of actions with a generic cost.
In the AMR example, the notion of \textit{duration}\footnote{Please note that \textit{time} is not considered explicitly. The durations are merely used to find the most preferred (fastest) way to restore safety.} is used.
If no value is specified, the planner will assume a value of~\textit{0}.
The planner only considers these values when choosing the best solution to restore safety.
For states where safety should only be maintained rather than restored, these values are ignored and can be all considered~\textit{0}.

Next, a few examples of actions of the AMR are given to provide further clarification.
\vspace{2mm}

\noindent
The action \textbf{stop\_conveyor}:
\begin{itemize}
    \item has a duration of \textit{1},
    \item controls one resource: \textit{conveyor},
    \item has one precondition: \(S_{Conveyor} = on\),
    \item has the nominal effect: \(S_{Conveyor} = off\),
    \item has no alternative effects.
\end{itemize}

\noindent
Since stopping the conveyor happens virtually instantly, the action gets assigned an arbitrarily low duration value.
Since there are no alternative effects specified, we assume that the action will never fail.\\

\noindent
The action \textbf{move\_to\_workstation\_1}:
\begin{itemize}
    \item has a duration of \textit{10},
    \item controls one resource: \textit{motors},
    \item has one precondition: \(S_{Location} = corridor\),
    \item has the nominal effect: \(S_{Location} = workstation\_1\),
    \item has one set of alternative effects: \(S_{Location} = corridor\).
\end{itemize}

\noindent
For moving to a specific location, we assign a relative duration of~\textit{10}.
The nominal effect is as expected, reaching \textit{workstation\_1}, and the alternative effect is expressed as the AMR not reaching the workstation.
In practice, this alternative effect can occur when this specific workstation is blocked or occupied.
This possibility is workstation-specific, and the effects can be modeled differently for every action.\\

\noindent
The action \textbf{receive\_workpiece}:
\begin{itemize}
    \item has a duration of \textit{3},
    \item controls the resources: \textit{conveyor} and \textit{motors},
    \item has the preconditions: \(S_{Location} = workstation\_1\) and \(S_{Load} = free\),
    \item has the nominal effects: \(S_{Conveyor} = on\) and \(S_{Load} = loaded\),
    \item has two sets of alternative effects:\\
    \(S_{Conveyor} = on\) and \(S_{Load} = free\), and\\
    \(S_{Conveyor} = off\) and \(S_{Load} = free\).
\end{itemize}

\noindent
A relative duration of \textit{3} is assigned to load transfer actions.
During the transfer of a workpiece, the motors are used to hold the AMR in place.
In this example, workpieces can only be obtained at \textit{workstation\_1},\linebreak hence the corresponding precondition.
For this action, two sets of alternative effects are considered.
The first one represents the outcome where a (for example, oddly shaped) workpiece gets stuck. The second one represents an even worse scenario where the entire conveyor gets blocked by receiving a (for example, too heavy) workpiece.

These three examples show how actions and their effects can be modeled.
If only one effect is given (the nominal effect), the outcome of the action is considered deterministic.
Alternatively, one or more alternative outcomes can be specified.
Recall that the alternative outcomes are not taken into account for the productive aspect of planning.
We do not (want to) rely on the alternative effects to reach a particular goal.
They are, however, considered when guaranteeing safety.

\subsection{Safety Rules}
Safety rules are used to add constraints to the behavior of the system to guarantee safe behavior.
In this section, we delve deeper into the use of reaction rules and introduce the new state rules.

\subsubsection{Reaction Rules}\label{reactionrules}
As explained in Section~\ref{TumatoRRules}, reaction rules are used to specify reactive behavior.
If a condition holds on the current state, an action is constrained to be executed or not.
We specified that if the AMR is present in the corridor with the conveyor on, the conveyor should be stopped:
\[(S_{Location} = corridor \wedge S_{Conveyor} = on) \Rightarrow Exec(\textbf{stop\_conveyor}).\]

However, we would prefer to guarantee that the conveyor is never on when the AMR is in the corridor (or at the charger\footnote{The previous reaction rule becomes: \((S_{Location} = corridor \vee S_{Location} = charger) \wedge (S_{Conveyor} = on) \Rightarrow Exec(\textbf{stop\_conveyor})\).}) in the first place.
Such behavior could lead to dropping workpieces and other dangerous situations for human agents in the factory.
Given our knowledge of the system, we specify:
\[(S_{Location} = corridor \vee S_{Location} = charger) \wedge (S_{Conveyor} = off)\]
\[\Rightarrow \neg Exec(\textbf{receive\_workpiece}) \wedge \neg Exec(\textbf{deliver\_workpiece}).\]
This rule has to be updated every time a new action is introduced that could turn on the conveyor.
We also have to specify:
\[(S_{Conveyor} = on) \Rightarrow \neg Exec(\textbf{move\_to\_\{x\}})\]
for every location \textit{x} for which moving toward \textit{x} could result in arriving at the corridor or charger.
This last expression can require a large number of reaction rules to be modeled, depending on the number of locations and how the effects of different move-actions connect them.
To considerably reduce the number of safety rules that have to be specified and hence, to make the specification less prone to errors, we introduce the concept of \textit{state rules} next.

\subsubsection{State Rules}\label{staterules}
This new kind of safety rule enables specifying constraints on reachable states rather than on the actions to execute.
We can simplify the previous example to one state rule:
\[(S_{Location} = corridor \vee S_{Location} = charger) \Rightarrow (S_{Conveyor} = off).\]

We introduce state rules in the form of \textit{desired safety conditions}.
The specified conditions should, according to the system's ability, always remain \textit{True}.
Alternatively, one can specify the unsafe conditions, which should be kept \textit{False}.
Translating between the two corresponds to negating the conditions.

The effects of an executed action should never lead to the violation of a state rule.
Formally, a state rule expresses a condition that must hold after executing any (set of) action(s) that the policy instructs for a state, regardless of which action effectuates first and regardless of which (nominal or alternative) effects occur.
Whether an action \textit{a} is allowed in a state \textit{s} can more formally be expressed as follows:
\[allowed(a,s) \Leftrightarrow \forall effect \in effect(a,s): effect \Rightarrow \{state\_rules\}\]
where \(effect(a,s)\) represents all possible effects (\(nominal \cup alternative\)) of action \(a\) in state \(s\) and \(effect \Rightarrow \{state\_rules\}\) denotes that the effect does adhere to all conditions described by the state rules.

If the current state was to violate a state rule, the next planned (set of) action(s) will always clear that violation (see Section~\ref{restoringSafety}).
Since the state rules do not allow any actions that could knowingly lead to undesired states, the system must have reached that state under the influence of an external force.
Further, if during planning for some state no instructions to restore safety exist, the planner notifies the user, indicating for which state no solution can be found.
When multiple such instructions are available, the planner selects that set of instructions that restores safety in the most preferred way.
For this example, we assigned durations to actions, hinting at the intention of restoring safety as quickly as possible.
Alternative approaches are to use costs and find the cheapest solution or to deal with risk explicitly.

\section{Discussion}\label{discussion}
In this section, we first discuss the modifications made to the set of constraints solved during planning.
We also present the findings from a preliminary experiment and analyze the resulting policy.
Finally, we discuss the challenges and potential limitations of the approach, hinting toward possible future work.

\subsection{Constraint-Based Planning}
The constraint-based approach providing the basis for this work has been introduced in the original work on Tumato~\cite{DinhHoangTung2017Sacr}.
The constraints that differ are the ones related to the new state rules.
While reaction rules could be incorporated directly into the constraint satisfaction problem as a constraint prohibiting or enforcing an action to be executed in a certain state, state rules require more insight.
For every execution step, the condition of the state rule is applied as a constraint to each state in the set of \textit{possible next states}.
State rules constrain every possible outcome of an action in the current state rather than the actions themselves.
This approach results in a number of new constraints, one for every possible outcome, enforcing a state rule conditionally to whether the action gets executed.
As a result, if an action \textit{could} violate one or more state rules, this action will be prohibited in the given state.
Practically, these constraints replace a (potentially large) number of constraints specified by reaction rules, as illustrated in Section~\ref{reactionrules}.
Complementary, state rules can \textit{enforce} the execution of actions if their effects are required to maintain safety.
Even if the current state is not safe, the next state is guaranteed to be safe, and the planner will minimize the required duration or cost to restore safety using a minimization objective.
Since the use of objectives is solver-specific, we refrain from elaborating further and only illustrate their potential use.
Finally, when no solution exists, the solver can refer to states for which no behavior can be generated to \textit{explain} why the specification is unrealizable, identical to the original feature of Tumato.

\subsection{Preliminary Experimental Results}
The example model described throughout this paper contains 32 states and 8 actions, and the planner takes between 2 and 3 seconds to obtain a policy on a \textit{1.6 GHz Dual-Core Intel Core i5}.
No significant memory usage was detected.
The initial CSP starts off with 518 constraints and ramps up to 5266 (as in Tumato, constraints are added automatically to obtain conflict-free plans to define a policy).
The number of decision variables starts at 823 and reaches up to 8487.
These numbers are slightly higher compared to using a reaction rule based approach but within the same order of magnitude.
For a scalability comparison, a somewhat larger model with 2688 valid states and 18 actions takes about 75 seconds, while memory usage remains negligible.
The CSP starts with 1130 constraints and ramps up to 5545.
The number of decision variables starts at 1787 and reaches up to 8911.

Practically, all the individual reaction rules have been replaced by the constraints generated from the state rule.
However, reaction rules require manual (more error-prone) implementation, and the state rules might cover situations the user did not anticipate.
Especially for more complex systems, this approach can be beneficial.
Finally, albeit more technical, we want to mention the impact on restoring safety \textit{the most preferred way} when the maximum planning length or number of possible unsafe states grows.
Now, the \textit{best} solution has to be found, rather than \textit{any} solution, as otherwise was sufficient.
Performance worsens because of the increasing number and complexity of the minimization objectives.

\subsection{The Policy}
Since the generated policies are complete, every valid state of the system has a corresponding entry in the policy.
In this section, a few well-chosen entries from the AMR's policy are presented.\footnote{A short clip of the AMR executing a policy generated by Tumato can be found \href{https://kuleuven-my.sharepoint.com/:v:/g/personal/jan_vermaelen_kuleuven_be/EQgdGpJ28eNCnuuap2i2mG4BOD2N92KTFuz1R013CdiIrA}{\textit{online}}.}

\begin{lstlisting}[basicstyle=\footnotesize]
  "S_Location": "corridor",                     "S_Location": "corridor",
  "S_Battery": "ok",                            "S_Battery": "low",
  "S_Load": "free",                             "S_Load": "free",
  "S_Conveyor": "off",                          "S_Conveyor": "off",
  "Actions": ["move_to_workstation_1"]          "Actions": ["move_to_charger"]
\end{lstlisting}
The first entry (left) instructs the AMR to move to workstation\_1, as it is currently not carrying a workpiece and there is sufficient battery power left.
The desired productive behavior appears.
In the second entry (right), as the battery level holds the value \textit{low}, a different (prioritized) conditional goal is active.
The AMR now moves to the charger.
In neither entry the state rule condition (specified in Section~\ref{staterules}) is violated, neither in the current state nor any foreseeable possible next state.

\begin{lstlisting}[basicstyle=\footnotesize]
  "S_Location": "workstation_2",                "S_Location": "corridor",
  "S_Battery": "ok",                            "S_Battery": "low",
  "S_Load": "free",                             "S_Load": "free",
  "S_Conveyor": "on",                           "S_Conveyor": "on",
  "Actions": ["stop_conveyor"]                  "Actions": ["stop_conveyor"]
\end{lstlisting}
In the next entry (left), the specified state rule is more prominent.
Without the state rule, the planner would instruct the AMR to start moving to \textit{workstation\_1}.
Since the conveyor will be active later on, turning it off now would be redundant.
It is clear that the state rule leads to executing \textit{stop\_conveyor}.
A naive implementation of state rules (or use of reaction rules) would allow simultaneous movement and stopping of the conveyor.
However, the adapted Tumato solver recognizes the non-deterministic order in which actions effectuate when multiple actions are executed simultaneously and ensures safety by permitting only the \textit{stop\_conveyor} action.
In the final presented entry (right), the current state violates the state rule.
To restore safety, the AMR is instructed to execute \textit{stop_conveyor}.
Although the more productive action \textit{move_to_charger} would also restore safety (as the only foreseeable next state is safe), the planner chooses the most preferred way to restore safety based on the durations of actions.

\subsection{Overview and Future Work}
Tumato succeeds in connecting \textit{theoretic} agents-based research, specifically on guaranteeing safe behavior, to the field of \textit{robotics}, where non-determinism is inevitable.
By including durations and alternative outcomes, we achieve a concise and expressive behavior specification.
State rules further enhance expressiveness and result in a stronger as well as less error-prone safety specification.
One main limitation of the approach is the use of discrete state variables.
For practical applications, a monitoring module has to be present to map the continuous world into a discrete state space.

Despite the promising combination of uncertainty and safety with the constraint-based planning approach, challenges remain open for further investigation.
The state rules currently enforce that any foreseeable outcome of the executed actions is safe.
When multiple sets of actions exist to \textit{restore} safety, the planner is capable of choosing the best one with regard to a value such as duration or cost.
When no actions are available to restore safety immediately, the planner will notify the user, indicating for which state safety can not be restored.
Although this is a desirable approach, future research could explore how allowing multiple successive (sets of) actions in unsafe states could be required to restore safety.
Furthermore, in practice, different safety constraints relate to different severities.
In the same way that actions can be assigned a specific value, the safety rules could be weighted to enable the constraint solver to find the overall best solution.
Finally, an empirical study will be conducted concerning the practical use of the adapted planner.
This study should consider the actual safety guarantees that are obtained as well as the specification effectiveness and comfort that is achieved.
Comparisons with the original version of Tumato, as well as with more traditional approaches, are in order.

\section{Conclusion}\label{conclusion}
This paper presents an approach to specify and generate safe and robust robot behavior.
For this purpose, we extend the existing constraint-based planning tool Tumato with the notion of uncertainty and state rules.
Robustness against uncertainty is achieved by extending actions with alternative, less desired and less likely, but foreseeable outcomes.
Further, state rules form a powerful approach for expressing safety rules based on state conditions rather than manually having to specify all situations that could lead to such unsafe states. 
This new approach requires an order of magnitude fewer rules to be specified and hence is less prone to errors.
Tumato translates the declarative specification, including the defined safety rules, automatically into a set of Constraint Satisfaction Problems.
Solving the CSPs yields an execution policy that inherently satisfies all the specified rules.
This approach enables the detection of unrealizable specifications early on.
The obtained policy is sound and complete by construction.

\paragraph{Acknowledgements} This research is partially funded by the Research Fund KU Leuven.

\nocite{*}
\bibliographystyle{eptcs}
\bibliography{generic}

\appendix
\section{The Specification}\label{appendix}
In this appendix, we use the AMR example to illustrate the extended format of specification supported by Tumato.

\definecolor{light-gray}{gray}{0.95}
\begin{lstlisting}[basicstyle=\footnotesize, backgroundcolor = \color{light-gray}]
BEGIN STATE VECTOR
state S_Location can be corridor, charger, workstation_1, workstation_2
state S_Battery can be low, ok
state S_Load can be loaded, free
state S_Conveyor can be on, off
END STATE VECTOR

BEGIN RESOURCES
resource MOTORS
resource CONVEYOR
END RESOURCES

BEGIN ACTIONS
action move_to_workstation_1
duration: 10
controlled resources: MOTORS
preconditions: S_Location is corridor
nominal effects: S_Location is workstation_1
alternative effects: S_Location is corridor

action move_to_workstation_2
duration: 10
controlled resources: MOTORS
preconditions: S_Location is corridor
nominal effects: S_Location is workstation_2

action move_to_charger
duration: 10
controlled resources: MOTORS
preconditions: S_Location is corridor
nominal effects: S_Location is charger
action move_to_corridor
duration: 2
controlled resources: MOTORS
preconditions: NOT S_Location is corridor
nominal effects: S_Location is corridor

action receive_workpiece
duration: 3
controlled resources: MOTORS, CONVEYOR
preconditions: S_Location is workstation_1, S_Load is free
nominal effects: S_Conveyor is on, S_Load is loaded
alternative effects: S_Conveyor is on, S_Load is free
alternative effects: S_Conveyor is off, S_Load is free

action deliver_workpiece
duration: 3
controlled resources: MOTORS, CONVEYOR
preconditions: S_Location is workstation_2, S_Load is loaded
nominal effects: S_Conveyor is on, S_Load is free
alternative effects: S_Conveyor is on, S_Load is loaded

action stop_conveyor
duration: 1
controlled resources: CONVEYOR
preconditions: S_Conveyor is on
nominal effects: S_Conveyor is off

action charge
duration: 50
controlled resources: MOTORS
preconditions: S_Location is charger
nominal effects: S_Battery is ok
END ACTIONS

BEGIN REACTION RULES // Please note, comments start with "//".
//rule: IF (S_Location is corridor OR S_Location is charger) AND S_Conveyor is on
//  THEN executing stop_conveyor
//rule: IF (S_Location is corridor OR S_Location is charger) AND S_Conveyor is off
//  THEN NOT executing receive_workpiece AND NOT executing deliver_workpiece
//  //AND NOT any future action that could turn the conveyor on
//rule: IF S_Conveyor is on THEN NOT executing move_to_corridor
//  AND NOT executing move_to_charger AND NOT executing move_to_workstation_1
//  //AND NOT any future action that could move the
//  //AMR away from NOT(corridor OR charger).
//And probably more rules
END REACTION RULES

BEGIN STATE RULES
rule: IF S_Location is corridor OR S_Location is charger THEN S_Conveyor is off
END STATE RULES

BEGIN GOALS
goal type: priority
when S_Battery is low then goal: S_Battery is ok
when S_Load is loaded then goal: S_Load is free
when S_Load is free then goal: S_Load is loaded
END GOALS

BEGIN CONFIG
max_plan_length: 5
END CONFIG
\end{lstlisting}

\end{document}